# Semantic Web Search based on Ontology Modeling using Protégé Reasoner

Monica Shekhar[1] and Saravanaguru RA. K.[2]

*School of Computing Science and Engineering, VIT University*

*Vellore, Tamil Nadu - 632014, India*

monicashekhar03@gmail.com, sarophd@gmail.com

**ABSTRACT**

**The Semantic Web works on the existing Web which presents the meaning of information as well-defined vocabularies understood by the people. Semantic Search, at the same time, works on improving the accuracy of a search by understanding the intent of the search and providing contextually relevant results. The paper describes a semantic approach towards web search through a PHP application. The goal was to parse through a user's browsing history and return semantically relevant web pages for the search query provided. The browser used for this purpose was Mozilla Firefox. The user's history was stored in a MySQL database, which, in turn, was accessed using PHP. The ontology, created from the browsing history, was then parsed for the entered search query and the corresponding results were returned to the user providing a semantically organized and relevant output.**

I. INTRODUCTION

Over the past years, the ever-increasing growth in the World Wide Web has brought into light the insufficiency of the currently existing techniques used for searching information on the web [2]. For a given query entered by the user, search for a target web page in most search engines is based on keyword-based searches and popularity based ranking. Although the results might be good enough, not all the search results turn out to be relevant to the given query. What is lacking in these search engines is a semantic structure, relationships between the information available over the web, making it difficult for the machine to understand the information asked for by the user and resulting in the loss of critical information while searching [3]. Therefore, semantic knowledge plays an important role.

The Semantic Web is, necessarily, a vision for the future of the World Wide Web where the available information is given a *meaning*, providing *logical connections of terms* and making it easier of the machine to integrate data and process the information available on the Web [8]. Semantics is the study of meaning. For the search query entered by the user, a Semantic Web Search ensures contextually relevant results by understanding *intent* and *meaning* of the query provided.

With the need for Semantic Web, W3C defined the first Resource Description Framework (RDF) specification for semantic interoperability in 1997. RDF required triple-based representations for Universal Resource Identifiers (URIs). Expression of structured vocabularies was then introduced in RDF Schema (RDFS). Web Ontology Language (OWL) provided greater expressivity in the objects and relations of the RDFS [4]. These ontologies provided a strong semantic structure to the data.

In the paper published by *Doms A. and Schroeder M.* in 2002, the first semantic search engine for biomedical texts using the Gene Ontology was published [5]. The Gene Ontology is a hierarchically structured vocabulary for molecular biology. *Sara Cohen Jonathan Mamou et al*, in 2003, presented a semantic search engine for XML called XSEarch [6]. It provided semantically related document fragments in accordance to the user's query. Later in 2004, *Li Ding Tim Finin et al* introduced a prototype Semantic Web search engine in a research project named *Swoogle* [7]. It is a search engine for the Semantic Web on the Web. *Swoogle* is a crawler-based indexing and retrieval system that searches for Semantic Web documents, instance data, terms, ontologies, etc. published on the Web.

These semantic search engines are designed to search for information in the World Wide Web. The objective of this research is to provide the user with semantically relevant results for the entered search query based on the *browsing history of the user*. The browser taken under study for this purpose was Mozilla Firefox. The browsing history of the user from the web browser was accessed and stored in MySQL database. The browsing domain considered in the research was "Apple". For all the URLs visited by the user under the considered domain, an ontology was created. For the query entered by the user, the created ontology was parsed and the visited URLs were displayed providing a semantically relevant list of URLs.

## II. SEMANTIC WEB LANGUAGE

The semantic web languages used in this application are RDF, RDFS and OWL [8]-[10]. The Resource Description Language is a general-purpose triple-based language used for representing information in the Web. The triples in RDF are represented as *subject-predicate-object*. RDF Schema is a semantic extension over RDF, providing vocabulary descriptions over the triples-based RDF. Web Ontology Language is used to make information available to be processed by applications, where the meaning of each term and their inter-relationships are explicitly represented. The representation of the terms and the relationship between thee terms is called an *ontology*. Due to the greater expressiveness of OWL, this language has the ability to represent machine interpretable content on the Web [8].

The various constructs in the OWL language, mostly used in the ontology created in the application, are the RDF Schema features (*rdf:subClassOf*, *rdf:Property*, etc.) for defining the classes, subclasses, properties, sub-properties, etc. , (In)Equality (*differentFrom, distinctMembers*, etc.) for specifying the inequalities between the various individuals, Property Restrictions (*Restriction*, *onProperty*, *allValuesFrom*, *someValuesFrom*, etc.) for defining restrictions on the individuals and Annotation Properties (*rdfs:label*, *rdfs:comment*, *AnnotationProperty*, etc.) for specifying the details for each individual [9].

## III. ARCHITECTURE

The research had been carried out in three modules:

- Web Content Extraction (Module 1)
- Semantic Knowledge Base (Module 2)
- Reasoner (Module 3)

The architecture of the system, in accordance to the modules, is shown below.

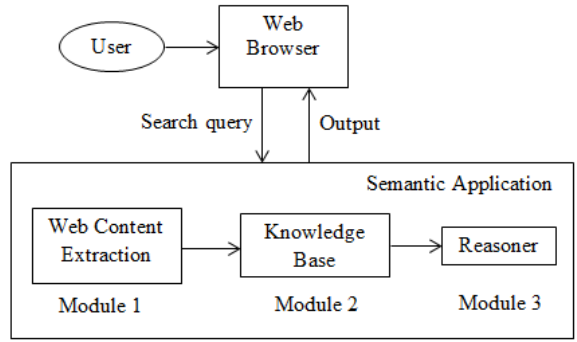

**Figure 1 Architecture of the system**

Here, the user enters the search query into the semantic application. The semantic application in itself consists of the three modules, wherein, the browsing history of the user was contained in module 1, the ontology created in accordance to the browsing history was contained in module 2 and module 3 parsed through the created ontology to return the result. This result was then displayed to user through the semantic application on the browser.

## IV. METHODOLOGY

### 1. Web Content Extraction (Module 1)

The coding for this module was done using python. Also, the browser under consideration was Mozilla Firefox. Mozilla Firefox stores all its data into a SQLite database. The following figure shows the SQLite Manager that was used to access the browsing history from the Mozilla Firefoz web browser.

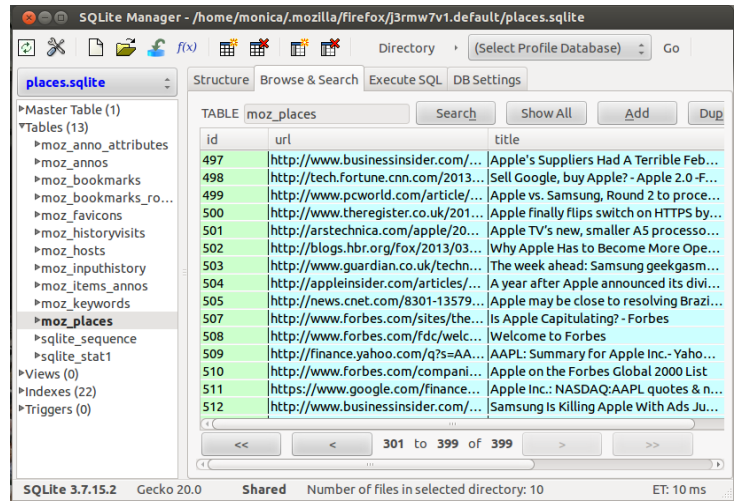

**Figure 2 User's browsing history in SQLite**

From the *moz.places* database, the user's browsing history was accessed and stored into a MySQL database using python, the python packages used being sqlite3 and MySQLdb. This is displayed in the figure as shown below.

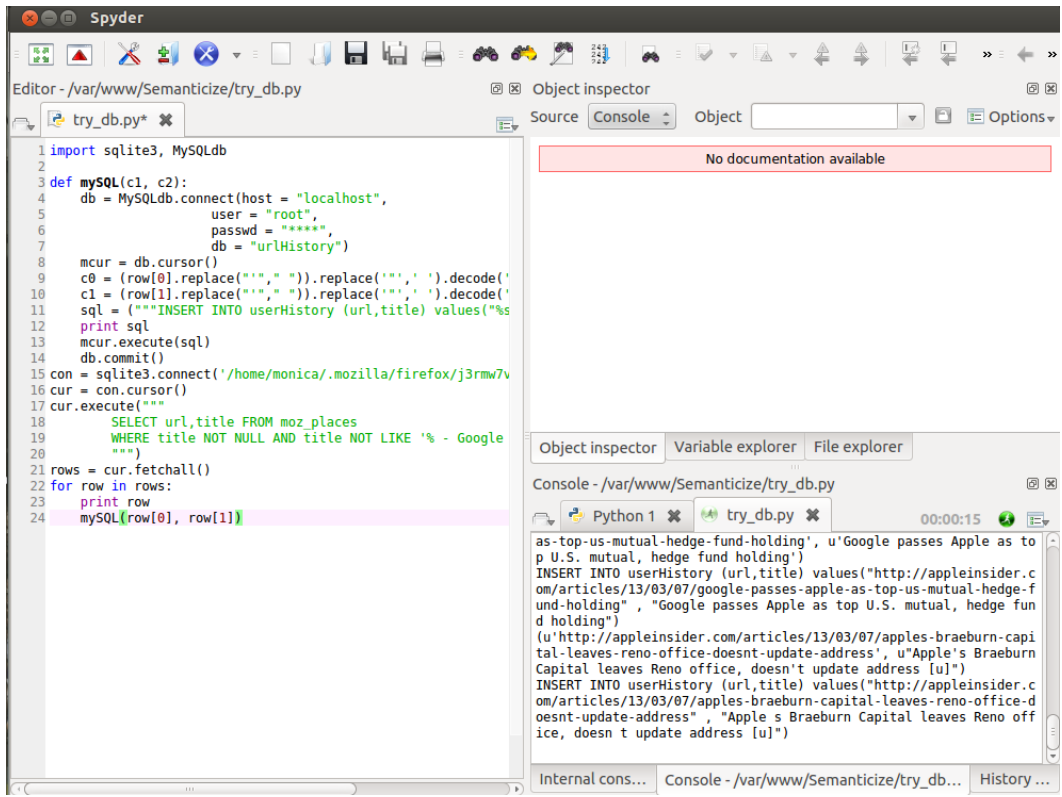

**Figure 3 Migration of data from SQLite to MySQL using python**

The above python code selects values from the sqlite3 database and stores it into a MySQL database, to be accessed in the PHP application. This python code was made to run in the background by adding a scheduler (the *sched* package) for the code to execute every 15mins. The web pages visited, in the meantime, would get updated into the MySQL table in intervals of 15mins.

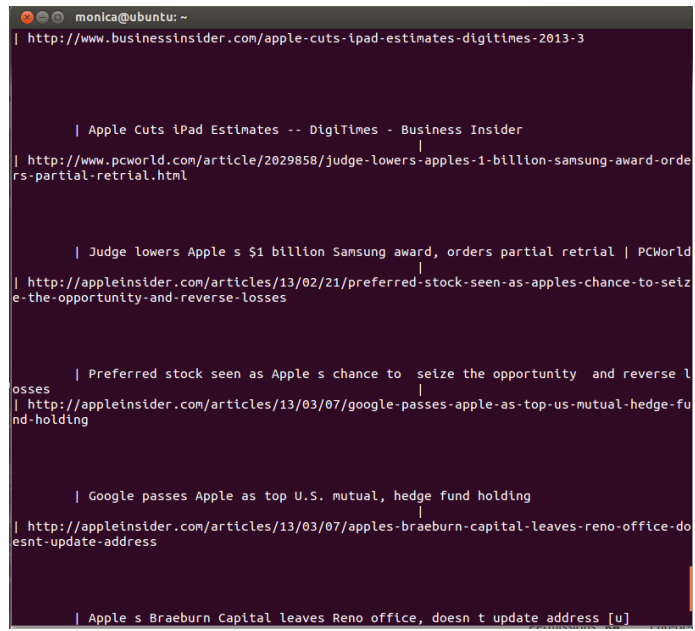

**Figure 4 Data displayed in MySQL**

Other python packages like BeautifulSoup4 and NLTK were used to extract the details of the visited URLs, like the title, description, etc. from the Web and were in turn stored into the MySQL database.

2.  **Semantic Knowledge Base (Module 2)**

The semantic knowledge base creation was done using the *Protégé-OWL editor*. Since the knowledge base domain, considered here, was "Apple", the ontology so created consisted of subclasses *AppleInc* (for the company) and *AppleFruit* (for the fruit). The various URLs obtained from Module 1 were added into the ontology as data properties for the corresponding classes, specifying the various schema features, OWL property restrictions, annotation properties, etc. [8][9]. Figure 5 shows the classes and the subclasses defined for the ontology, Figure 6 shows the various data properties in the ontology and Figure 7 shows the individuals created in the ontology.

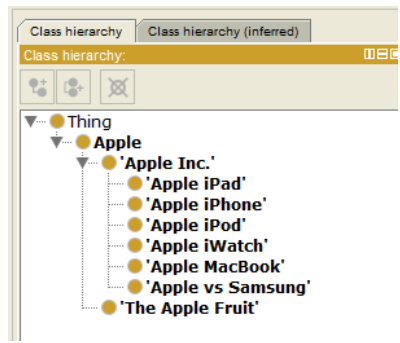

**Figure 5 Class hierarchies for the keyword 'Apple' in Protégé**

**Figure 6 Data properties in the ontology**

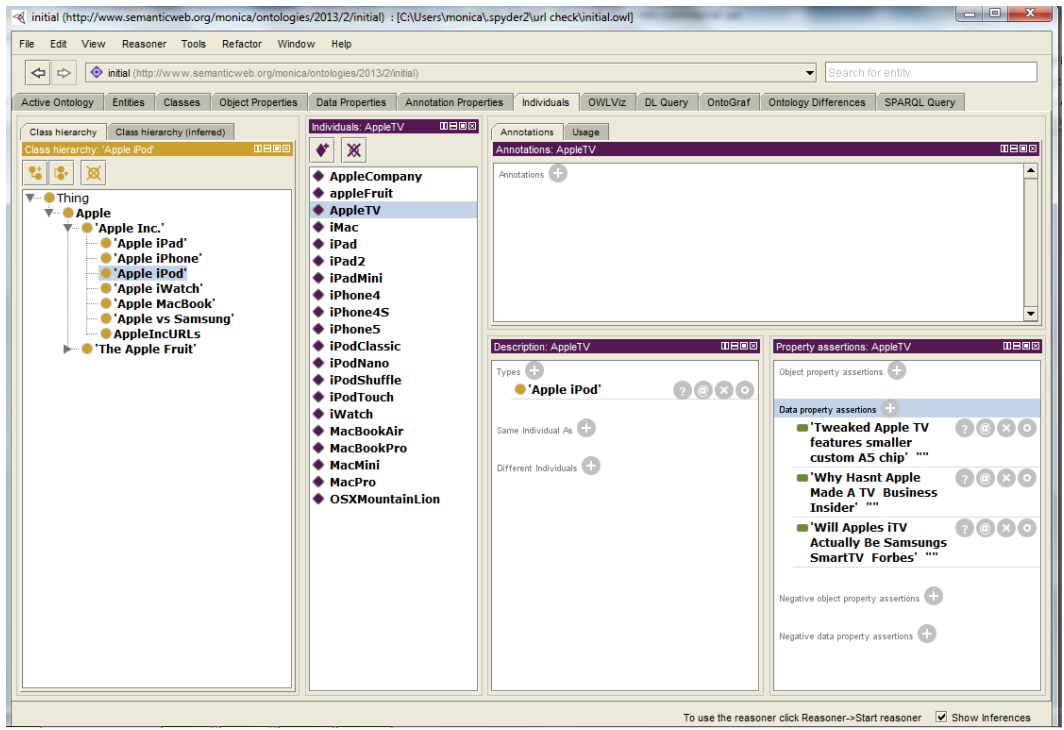

**Figure 7 Individuals to the classes**

The property restriction *allValuesFrom* was used to differentiate between the URLs of the two Apple subclasses, where the subclasses *AppleInc* and *AppleFruit* were defined to be disjoint with each other. The various restrictions applied on the classes in the ontology is shown in the figure below.

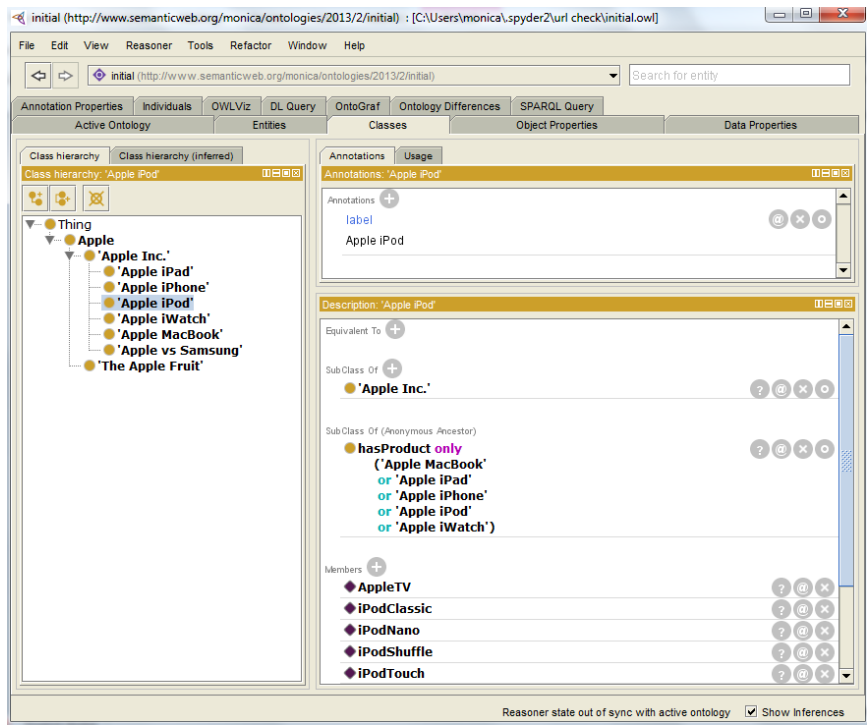

**Figure 8 Restrictions applied on the classes**

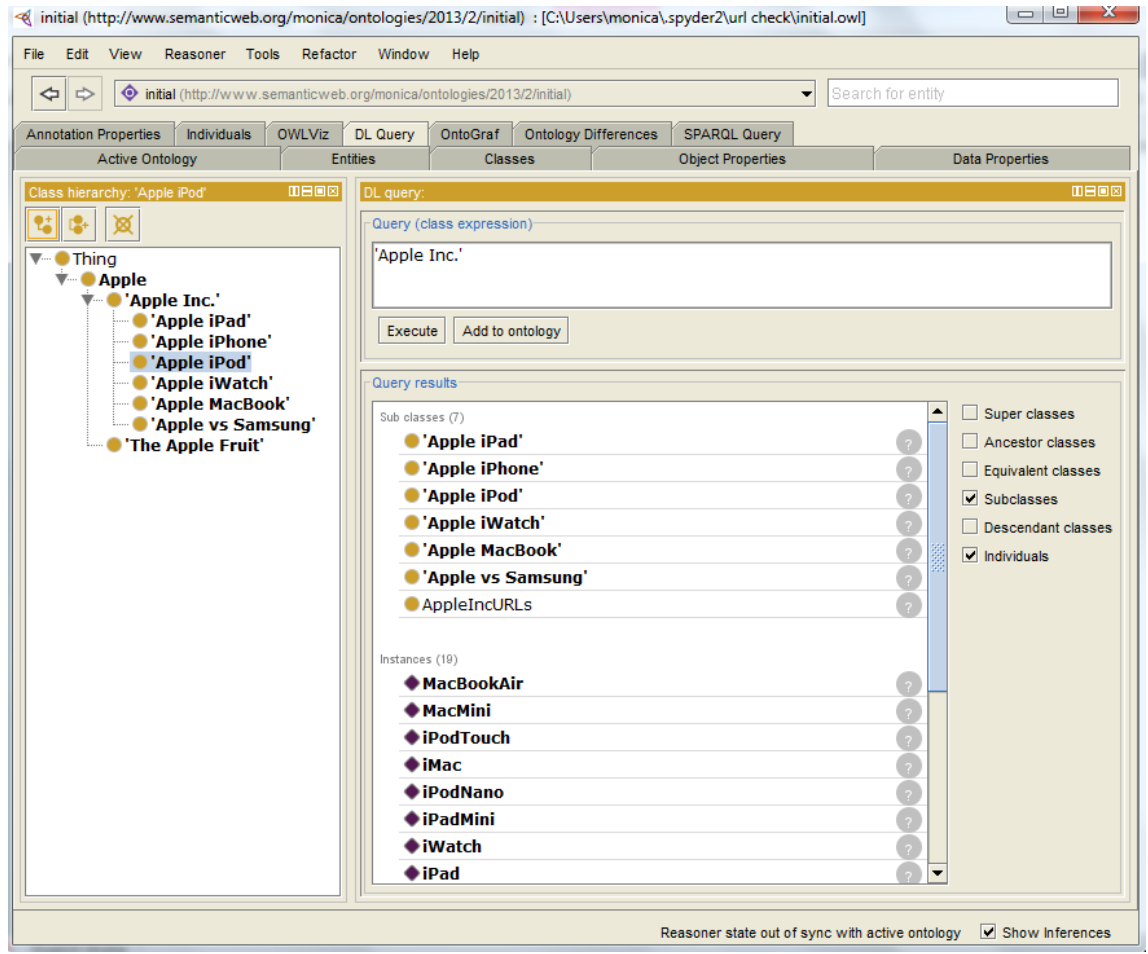

Figure 9 DL Query displaying the subclasses, descendant classes and individuals of a keyword – Apple Inc.

### 3. Reasoner (Module 3)

The final module in the application brings in both Module 1 and Module 2 to work in the backend. The Reasoner acts as the frontend of the application where the user enters the search query and the results were displayed. This module is coded using PHP, HTML5 and CSS. Figure 10 shows the frontend as visible to the user.

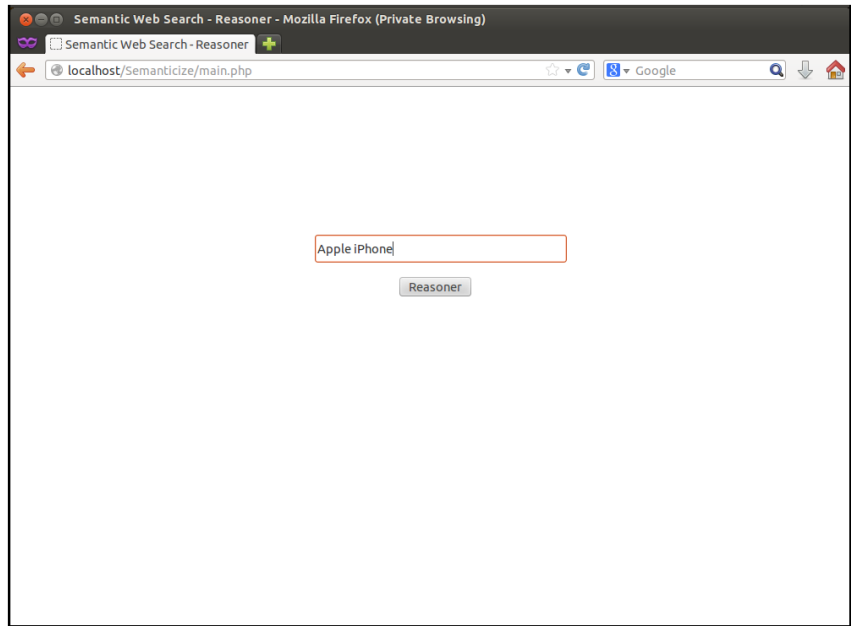

Figure 10 Search query interface

The ARC2 library available in PHP was used for parsing through the OWL file created in the previous module. ARC2 is a flexible RDF system for semantic web and PHP. It provides SPARQL and easy RDF parsing for LAMP systems. The ARC2 library was made available in the PHP module, which was then used to query the ontology using SPARQL.

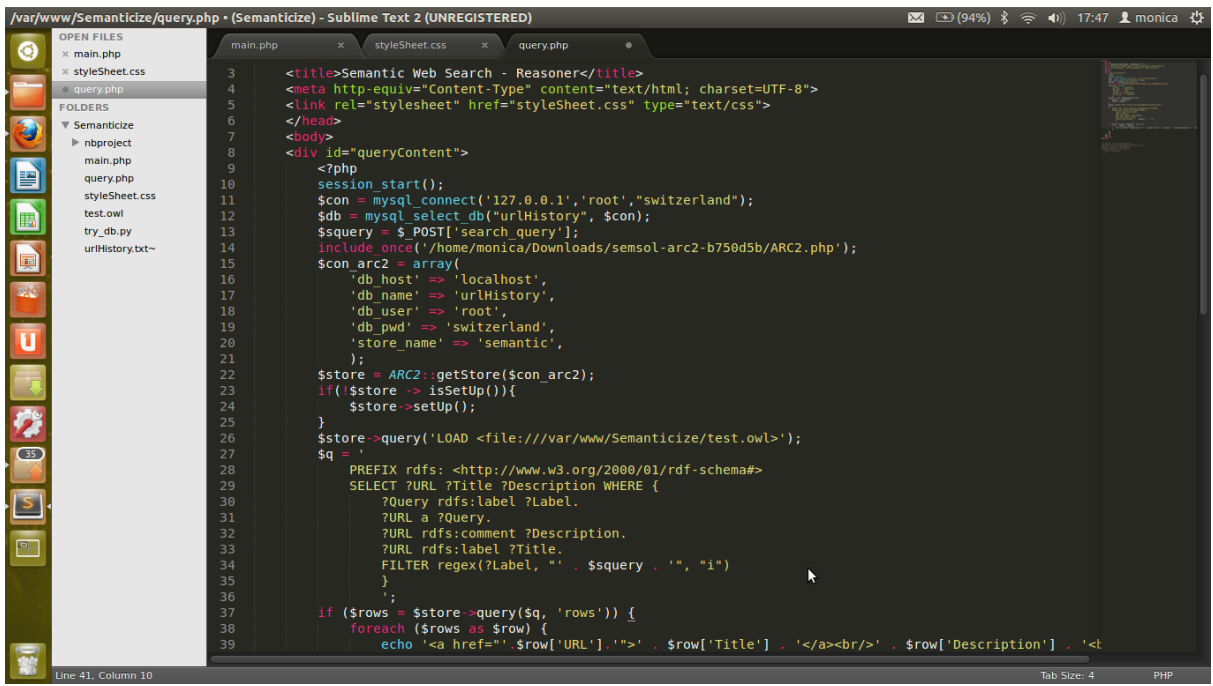

Figure 11 Querying the ontology using ARC2 and SPARQL

So, the ontology, saved as an OWL file, was accessed, parsed and queried in the PHP application, where the query entered by the user was passed as the query to the file. Here, SPARQL is used to query the OWL file, which returns the results from the created ontology. These results are then displayed to the user in the output page. HTML5 and CSS are used for designing the application to make it look more user-friendly. This user interface is displayed in Figure 12. Also, to provide more semantically relevant results, the information related to the entered query can also be displayed along with the URLs.

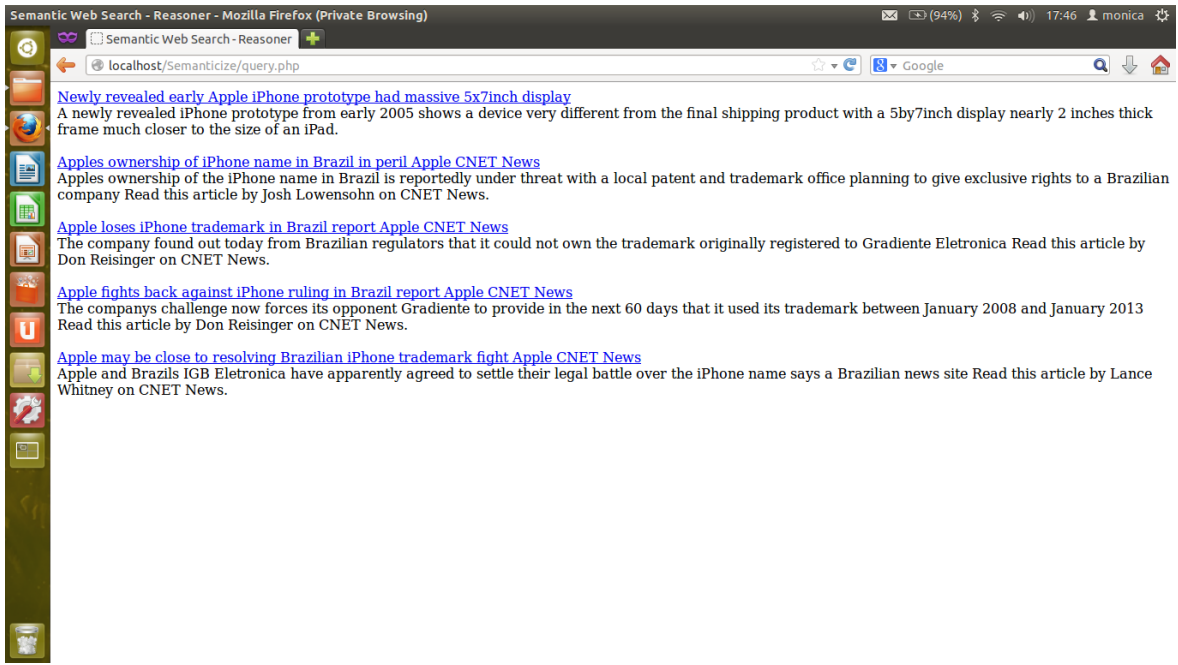

Figure 12 List of visited URLs semantically arranged

## V. RESULT AND FURTHER WORK

The application, hence, aides the search of the user by providing a list of useful and relevant web pages earlier visited by him, in accordance to the query, thereby providing him with results that would be relevant and helpful.

The major area of further work in the system would be to make the application work for web browsers other than Mozilla Firefox. Accessing databases of Google Chrome, Safari, Internet Explorer, etc. and storing them in the MySQL database along with the existing database would make the application work for the other web browsers as well. To make the application more accessible to the user, the system can be made available as a plugin to be installed into the web browser, providing semantically organized results to the user simultaneously with the current searches being made.

Also, further work can be done in increasing the knowledge base by providing automation in the process of adding information to the ontology.

## VI. CONCLUSION

With the increasing amount of data being stored on the Web every day, Semantic Search would ensure the provision of useful and relevant information to the user. This application works on making the procedure more user-friendly by providing contextually relevant visited pages to the user along with the searches available to the user otherwise through the various search engines. Understanding the user's query and providing just the content required is the objective of the application.